\documentclass[aps,prd,reprint,nofootinbib,superscriptaddress]{revtex4-2}
\usepackage[english]{babel}
\usepackage[utf8]{inputenc}
\usepackage[colorinlistoftodos, color=green!40, prependcaption]{todonotes}
\usepackage{orcidlink}
\usepackage{amsthm}
\usepackage{mathtools}
\usepackage{physics}
\usepackage{xcolor}
\usepackage{graphicx}
\usepackage{adjustbox}
\usepackage{placeins}
\usepackage[T1]{fontenc}
\usepackage{lipsum}
\usepackage{csquotes}
\usepackage{hyperref}
\usepackage{lipsum}
\usepackage[caption=false]{subfig}
\begin{document}
\title{LIQUIDating the Gallium Anomaly}
\author{Garv Chauhan\,\orcidlink{0000-0002-8129-8034}}
    \affiliation{Department of Physics, Arizona State University, 450 E. Tyler Mall, Tempe, AZ 85287-1504 USA}
    \affiliation{Center for Neutrino Physics, Department of Physics, Virginia Tech, Blacksburg, VA 24061, USA}
\author{Patrick Huber\,\orcidlink{0000-0002-2622-3953}}
    \affiliation{Center for Neutrino Physics, Department of Physics, Virginia Tech, Blacksburg, VA 24061, USA}    
\date{\today}

\begin{abstract}
The gallium anomaly has a global significance of greater than $5\sigma$. Most viable BSM solutions quickly run into strong tensions with reactor and solar neutrino data. We propose to use indium (${}^{115}\text{In}$) as a target as it offers a low threshold and reasonably high cross section.  The neutrino-indium charged current cross section can be calibrated using the well-constrained solar ${}^{7}\text{Be}$ neutrino flux that lies very close in energy to the ${}^{51}\text{Cr}$ neutrino lines. The triple coincidence provided by ${}^{115}\text{In}$ neutrino capture can be fully exploited by an opaque scintillation detector that also provides energy and position information. We show that a $100$\,ton indium target combined with 2 source runs of a $3.4$\,MCi ${}^{51}\text{Cr}$ source can probe the complete parameter space of the gallium anomaly, both in the context of a vanilla sterile neutrino as well as more involved BSM scenarios. 

\end{abstract}

\maketitle

\textit{\textbf{Introduction}} The deficit of event rate induced by mono-energetic neutrinos produced from intense radioactive sources, in gallium-based radiochemical measurements, is commonly termed as the \textit{Gallium Anomaly}~\cite{SAGE:1998fvr,GALLEX:1997lja,Kaether:2010ag,Barinov:2022wfh}. The underlying charged current process is given by
\begin{equation}
    \nu_e + {}^{71}\mathrm{Ga} \rightarrow e^- + {}^{71}\mathrm{Ge}\,.
\end{equation}
The anomaly currently stands at a global significance of greater than $5\sigma$~\cite{Berryman:2021yan}. While there can be possible sources of errors from standard physics, no single factor can explain the anomaly. A simple sterile neutrino explanation runs into strong tension with current solar neutrino bounds, reactor neutrino measurements and KATRIN data~\cite{Berryman:2021yan,Giunti:2022btk, KATRIN:2025lph}. 

The possible standard resolutions include possible errors in source strength determination, neutrino capture cross-section on germanium and radiochemical detection efficiency~\cite{Huber:2022osv,Brdar:2023cms}. While a combination of these explanations could explain away the anomaly, any one of these reasons by themselves is not sufficient. This might be a clue for the existence of new physics. However, since the solar and reactor neutrino energies span the same energies as the $^{51}$Cr and $^{38}$Ar neutrino sources, many elegant solutions quickly run into tension with the solar and reactor data. Some of these explanations which primarily rely on light sterile neutrinos ($\nu_s$) includes vanilla eV-scale $\nu_s$, decaying $\nu_s$, $\nu_s$ with CPT violation and $\nu_s$ coupled to a ultralight scalar~\cite{Brdar:2017kbt,Giunti:2022btk,Barinov:2022wfh,Giunti:2010zs,Davoudiasl:2023uiq}

The gallium data to-date has been obtained in radio-chemical experiments and hence no information on the event energy or baseline dependence (apart from the 2-zone BEST measurement) is available. We propose a concept where both the baseline and energy can be measured in real-time with high accuracy by combining the LENS idea to use $^{115}$In as target~\cite{Grieb:2006mp} and the LiquidO detector concept that allows for good spatial and energy resolution as well as a high indium loading of the scintillator~\cite{LiquidO:2019mxd}. The neutrino source will still be a $^{51}$Cr source\footnote{The only solar neutrino flux at the $^{51}$Cr energy is the CNO flux, but Borexino has insuficcient statistics to effectively test the anomaly, for details see the App.~\ref{sec:Borexino}.}, but now the low-energy line at 426\,keV can be used as a source strength calibration. Borexino~\cite{Bellini:2011rx} has measured the $^7$Be neutrino flux at earth with a precision of 2.6\%, but the $^7$Be solar neutrino  line is not at the right energy to probe the gallium anomaly itself. However, it is very close in energy to the ${}^{51}\text{Cr}$ neutrino lines and hence can be used to determine the $\nu_e-^{115}$In cross section. Borexino uses elastic neutrino-electron scattering for its measurement, a purely leptonic process that can be computed precisely from first principles. As a consequence \emph{all} ingredients can be constrained directly from \emph{data} or well-understood Standard Model physics. Such a measurement, therefore will close the window on all SM explanations and any deficit will be an unequivocal sign of new physics. The information about baseline and energy dependence effectively will  pin-down the underlying physics. This can be achieved, as we will show, with 2 runs of a 3.4\,MCi $^{51}$Cr source and a 100\,ton indium target in a LiquidO-style detector.

In this work, we specifically investigate a class of BSM solutions (proposed in Ref.~\cite{Brdar:2023cms}) that are still viable for the resolution of this anomaly. This solution primarily relies on engineering a fine-tuned MSW/parametric resonance for $\nu_s$ through its interactions with ultralight dark matter or dark energy. Such narrow resonances are placed directly at the $^{51}$Cr neutrino energies and appearing for only very short-baselines of about $1$-$2$\,m, hence alleviating any possible tensions from measurement of $pp$ and ${}^7$Be solar neutrino fluxes.

\begin{figure}[t]
    \centering
    \includegraphics[width=\columnwidth]{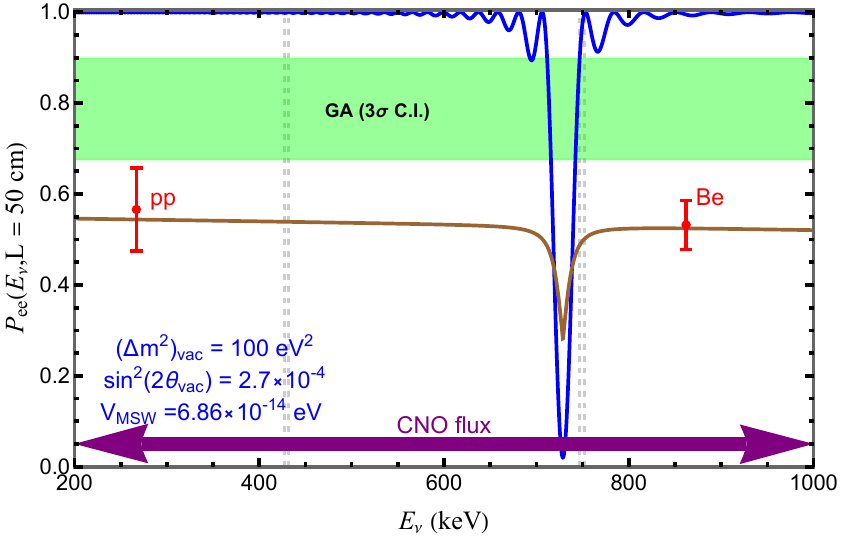}
    \caption{Electron Survival Probability $P_{ee}$ with $\nu_s$-DM MSW resonance
    for a sample point in $95\%$ C.I. allowed region from all GA experiments (see Fig.~\ref{fig:gridVacEff}).}
    \label{fig:p2}
\end{figure}

\begin{figure*}[t!]
    \centering
    {\includegraphics[width=0.49\linewidth]{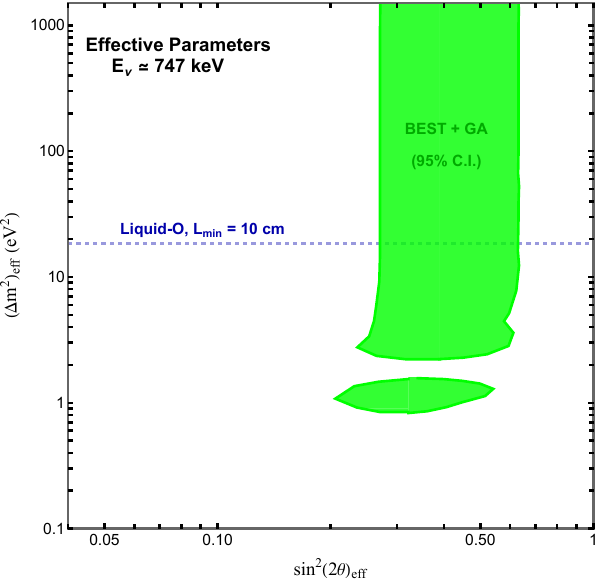}}
    {\includegraphics[width=0.49\linewidth]{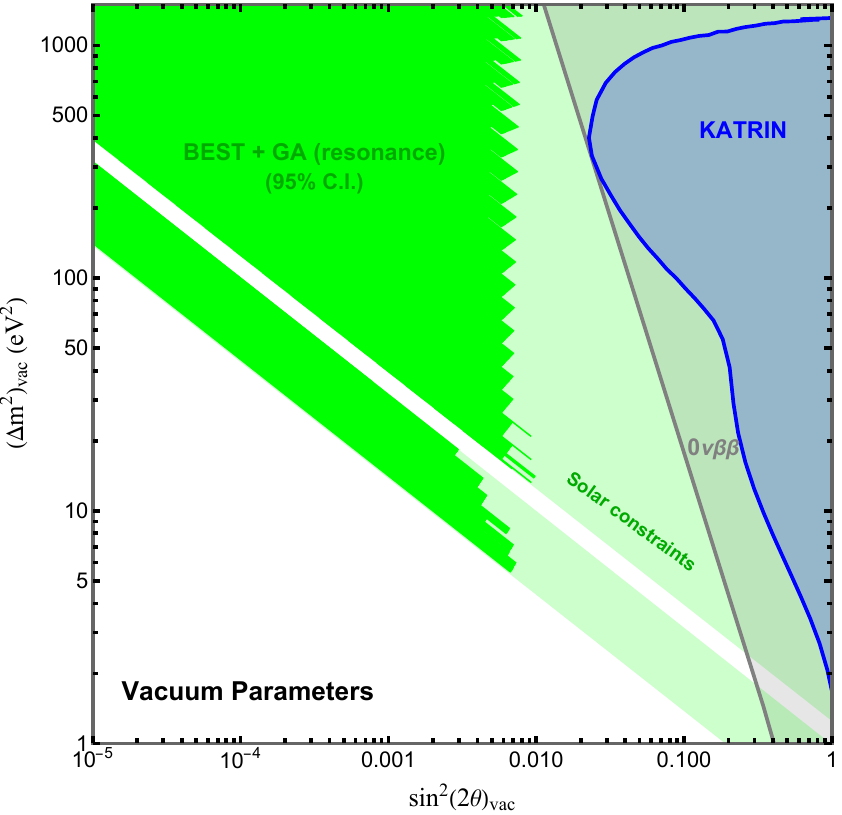}}
    \caption{The  best-fit and exclusion contours ($95\%$ C.I.) for the Gallium Anomaly in effective and vacuum parameter space for $\nu_s$-DM MSW resonance model.}
    \label{fig:gridVacEff}
\end{figure*}

\label{sec:res}
\textit{\textbf{MSW Resonance}} In general, BSM model construction to explain GA anomaly has been challenging, with severe tensions arising from solar, reactor, accelerator and cosmological measurements. Since vanilla $\nu_S$ solution quickly runs into such issues, we focus on a new class of model utilizing tuned MSW or parametric resonance induced for sterile neutrinos through their interactions with ultralight dark matter or dark energy~\cite{Brdar:2023cms}. Such tuned resonant conversion for $\nu_e\rightarrow\nu_s$ at energies close to dominant $^{51}$Cr neutrino line can explain GA. We specifically focus on the scenario of ultralight vector dark matter sourcing a potential for $\nu_s$ leading to a MSW resonance. Since this case leads to a suppressed effect on the solar neutrino flux compared to the parametric resonance case, this makes our model choice to be conservative. This model  can easily evade the solar constraints since the impact is too narrow in energy to be effectively constrained (see appendix). The sign of the MSW potential ensures that the resonance lies only in the $\nu$-sector, while being absent for $\bar{\nu}$'s. Thereby, evading $\bar{\nu}$ measurement constraints from reactor and accelerator experiments. The resonance is chosen to lie around the dominant $^{51}$Cr neutrino line $\sim 747$ keV. Therefore, any constraints from high-energy accelerator neutrino measurements ($\gg 747$ keV) can be easily avoided. Finally the kinematical bounds from KATRIN or $0\nu\beta\beta$ experiments, can be avoided by choosing a small-enough mixing angles for a given vacuum $\nu_s$ mass, while suitably choosing a MSW potential $V$ to explain the GA.  Therefore, this solution ensures that observable effects only occur for neutrinos at the $^{51}$Cr energy (for e.g. see Fig.~\ref{fig:p2}).

The MSW resonance condition from the presence of ultralight species modifies the probability for detection of electron neutrino for shorter baselines in the following way~\cite{Brdar:2023cms}
\begin{equation}
    P_{ee}^{\text{MSW}}(E_\nu,L)= 1- \sin^2{2\theta^{\text{eff}}_{e4}}\sin^2{\frac{\Delta m^2_{\text{eff}} L}{4 E_\nu}}
\end{equation}
where $\Delta m^2_{\text{eff}}=\Delta m^2(\sin{2\theta^{\text{vac}}_{e4}}/\sin{2\theta^{\text{eff}}_{e4}})$, where $\Delta m^2$ is the vacuum $\nu_s$ mass,  $\sin{2\theta^{\text{vac}}_{e4}}$ is the vacuum mixing angle and the mixing angle in matter $\sin{2\theta^{\text{eff}}_{e4}}$ is determined by the interactions of the $\nu_s$ with the ultralight dark matter or dark energy,
\begin{equation}
    \sin{2\theta^{\text{eff}}_{e4}} =\frac{\frac{\Delta m^2_{\text{vac}}}{2 E_\nu}\sin{2\theta^{\text{vac}}_{e4}}}{\sqrt{\left(V-\frac{\Delta m^2_{\text{vac}}}{2 E_\nu}\cos{2\theta^{\text{vac}}_{e4}}\right)^2+\left( \frac{\Delta m^2_{\text{vac}}}{2 E_\nu}\sin{2\theta^{\text{vac}}_{e4}}\right)^2}} 
    \label{eq:sin2eff}
\end{equation}
where $V$ is the matter potential difference between $\nu_e$-$\nu_s$. We show the $95\%$ C.I. allowed regions from all GA experiments (SAGE~\cite{SAGE:1998fvr}, GALLEX~\cite{GALLEX:1997lja,Kaether:2010ag}, BEST~\cite{Barinov:2022wfh}) for $\nu_s$-DM MSW resonance model in effective and vacuum parameters space in Fig.~\ref{fig:gridVacEff}. The effective parameter values have been calculated for $E_\nu\sim 747$ keV. The vacuum parameters for a given set of effective parameters have been determined using definition of $\Delta m^2_{\text{eff}}$ and Eq.~\eqref{eq:sin2eff}. Note that MSW potential $V$ for a given vacuum parameter has not been shown explicitly to enhance clarity, a given set of vacuum parameters ($\Delta m^2,\sin{2\theta^{\text{vac}}_{e4}}$) can correspond to multiple values of $V$ that can explain GA. We also note that the best-fit allowed regions for vanilla eV-scale sterile neutrino~\cite{Barinov:2022wfh} differ only slightly for our effective parameters best-fit at $E_\nu\simeq 747$ keV for MSW resonance case, since due to the absence of any oscillation around the $427$ keV $^{51}$Cr neutrino line, the required mixing angles to explain the lower Gallium experimental rates are relatively higher. Therefore, the latter best-fit plots shifts slightly to the right. 

\label{sec:IndDet}
\textit{\textbf{Indium Detector}}
In 1976, Raghavan~\cite{Raghavan:1976yc,Raghavan:1997ad,Raghavan:2001jj} put forward a novel proposal to detect the low energy pp-solar neutrinos through charged-current driven neutrino capture process on Indium (In),  
\begin{equation}
    \nu_e + {}^{115}\text{In} \rightarrow  {}^{115}\text{Sn}^* + e^{-}
\end{equation}
with a reaction threshold $Q=114$ keV.  The captured neutrino energy $E_\nu$ can be reconstructed directly from the final state electron energy $E_e=E_\nu-Q$.  The ${}^{115}\text{Sn}^*$ state undergoes delayed de-excitation by emitting two photons of energies $116$ keV and $498$ keV.
\begin{equation}
    {}^{115}\text{Sn}^* \rightarrow  {}^{115}\text{Sn} + 2\,\gamma
\end{equation}
For future studies, we provide the neutrino capture cross-section on ${}^{115}\text{In}$ ($\sigma_\text{In}(E_{\nu})$) in appendix~\ref{app:crosssection}.

The originally suggested idea was to dissolve 5-10 wt\%  indium in liquid organic scintillator~\cite{Raghavan:1976yc,Pfeiffer:1978zz,Barabanov:2023dfl}. More recently opaque liquid scintillator, that has a shorter scattering than transmission length, has been considered for this application~\cite{LiquidO:2019mxd}. This localizes the light without segmentation and by using wavelength-shifting optical fibers the localized light can be transported over practically arbitrary distances. In such a scheme a very high indium concentration is possible since light transport happens in the fibers and not in the primary scintillator. This is known as the LiquidO concept~\cite{LiquidO:2019mxd} and is a promising technology to use ton-scale indium targets. In the following we will specify the amount of indium instead of the total detector mass since the actually achievable indium concentration is not yet determined. 
The lower-energy gamma of the triple coincidence is with 116\,keV not an easy target for a detector design with limited photon collection efficiency, achieving a reasonable detection probability for those event probably requires between 10-25 detected photons, or about 100-250 detected photons per MeV of deposited energy. This translates into an energy resolution at the $^{51}$Cr line of about $80-130$\,keV. Initial prototypes of the LiquidO technology show promising results~\cite{APILLUELO2025170075} and we assume that the required levels of light collection will eventually be reached~\cite{LiquidO:2019mxd}. 
We furthermore assume that no backgrounds are present, which is plausible since the triple coincidence of the indium reaction precludes accidental backgrounds. The same applies for any backgrounds originating from the $^{51}$Cr source: the backgrounds from the chromium source are entirely due to impurities in the chromium feed material that get neutron-activated and produce high energy gammas~\cite{Gavrin:2021lam} but their rate will be suppressed by the shield to below the one from ${}^{115}\text{In}$ decay. 

\textit{\textbf{Testing the Resonance}}
\begin{figure*}
    \centering
    \includegraphics[width=0.495\linewidth]{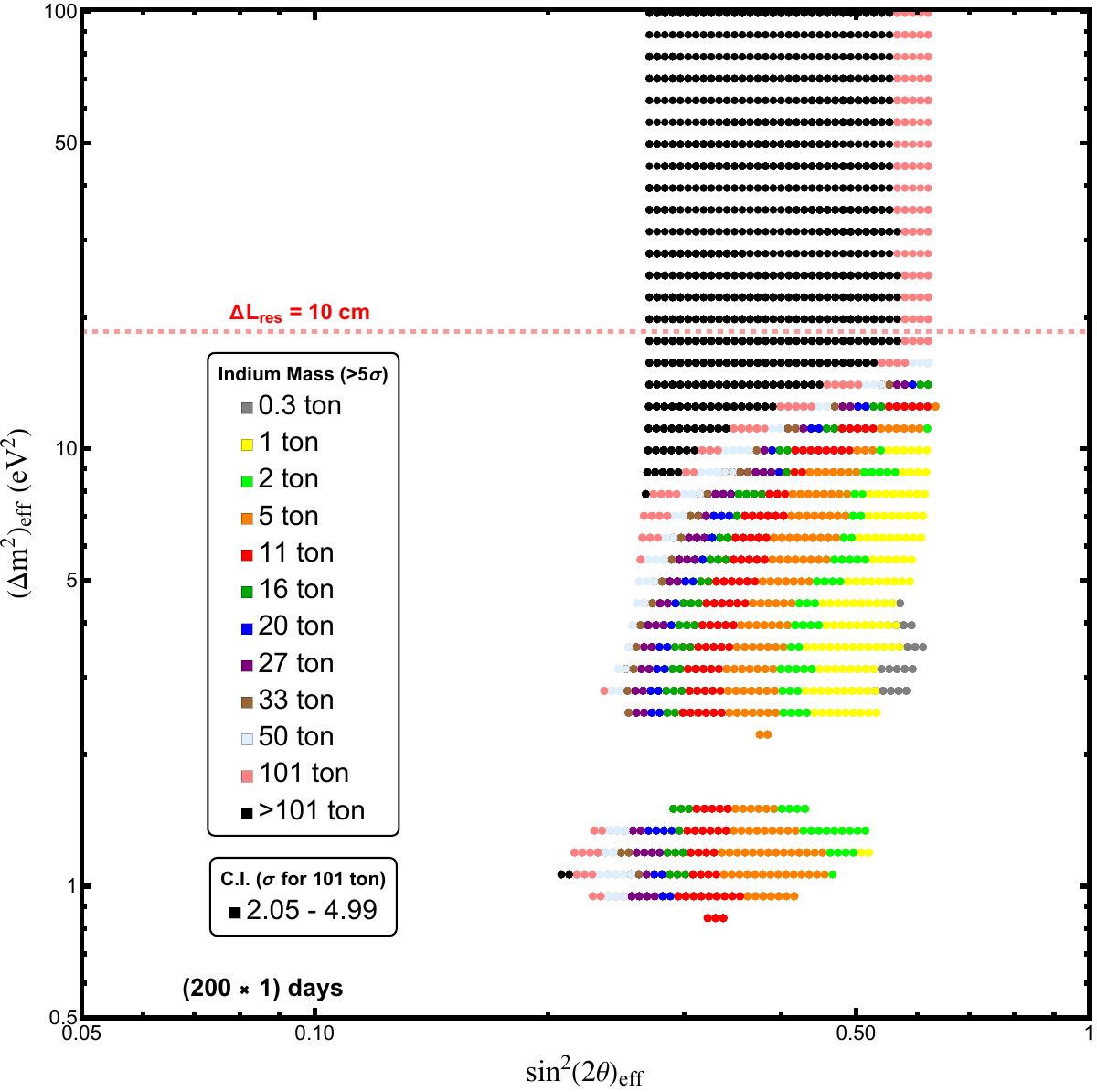}
    \includegraphics[width=0.495\linewidth]{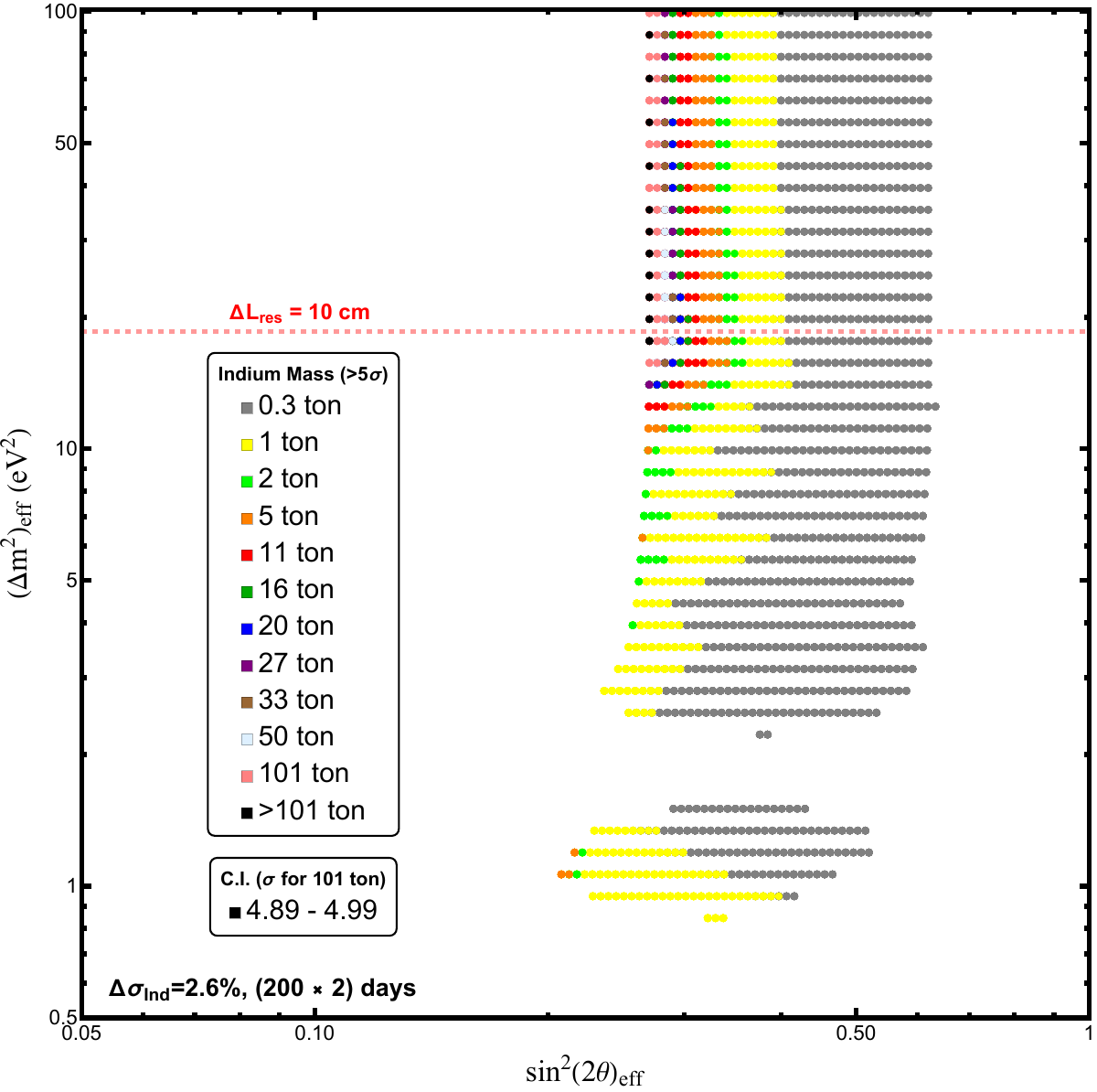}
    \caption{Plots for sensitivity of the In-doped LiquidO setup. The colored points indicate the required Indium mass to reach confidence interval (C.I.) of greater than $5\sigma$. All the points shown here are sampled uniformly from the allowed parameter space. The left panel corresponds to single source run with a floating precision on the indium neutrino capture cross-section, and a double source run with $\sigma_{\sigma\text{In}}=2.6\%$  in the right panel.}
    \label{fig:gridReach}
\end{figure*}
We will study the sensitivity of the In-doped LiquidO setup to the entire $95\%$ C.I. best-fit allowed region for the $\nu_s$-DM MSW resonance model, to explain GA. We assume a 3.4\,MCi Cr source placed at the center of the detector setup to obtain a high intensity neutrino flux. We estimate the physical size of the neutrino emitting region to be about 10 cm~\cite{Gavrin:2021lam}, which also is the limit on the shortest wavelength of oscillation that can be resolved. The overall size including the radiation shielding is about $\sim40$\,cm, which limits how close to the source we can have active detector mass and thus sharply affects the event rate.
Note, that in this geometry, the event rate scales with the detector radius and hence only with $\text{M}_{\text{Ind}}^{1/3}$, {\it i.e.} to double the event rate eight times more indium mass ($\text{M}_{\text{Ind}}$) is required. The density of indium in the detector medium also matters since it determines how much indium is within a given distance from the source; we assume 20\% indium loading.
As expected, for the detection of solar $^7$Be neutrinos, the event rate is as usual proportional to the total indium mass. Despite the significant thickness of the shielding, there will be residual gamma radiation from neutron-activated trace impurities contained in the source material~\cite{Gavrin:2021lam,Huber:2022osv}, however the $^{115}$In triple coincidence will effectively eliminate this background in particular thanks to the excellent position resolution of the LiquidO technology : high-energy gammas will Compton scatter multiple times and therefore can be clearly identified. The energies and relative intensities of the four neutrino lines in ${^{51}}\text{Cr}$ decay are $747$ keV ($81.63\%$), $752$ keV ($8.49\%$), $427$ keV ($8.95\%$) and $732$ keV ($0.93\%$)~\cite{Acero:2007su}. Since in the $\nu_s$-MSW resonance model, we expect a strong suppression for $P_{ee}$ at dominant Cr neutrino line, the difference in the relative intensities and wide energy separation between the $\sim750$ keV and $\sim430$ keV lines play a crucial role in our analysis. The details of the cross section, event rate and $\chi^2$-calculation can be found in the appendix. 

The results in Fig.~\ref{fig:gridReach} are reported in terms of effective oscillation parameters defined at $747$ keV. For the left-hand panel we assume that there is no external constraint on the $\nu_e-^{115}$In cross section, that is the rate is normalized solely by the 427\,keV data. For the right-hand figure we assume that enough events from $^7$Be solar neutrinos have been observed to obtain a statistical uncertainty below 2.6\% (the error of the Borexino measurement) and hence the cross section is constrained to within this value. To reach a statistical precision at this level requires approximately 100\,tons of indium target exposed for 10 years. If another measurement of the solar $^7$Be flux with higher precision became available this experiment would become much easier and significantly smaller indium mass would be required, such improvements may come from large liquid noble gas direct dark matter detection experiments~\cite{McDonald:2024osu,Aalbers:2022dzr}. We  conclude based in the Borexino measurement that even ton-scale experiments can provide useful constraints (at $>3.5$-$\sigma$ level for the entire parameter space) and that a 100\,ton experiment can test all of the parameter space at more than 5-$\sigma$ level.  None of the proposed  measurements rely on  radiochemistry or on nuclear matrix elements for inverse beta decay on nuclei.

\label{sec:conclusion}
\textbf{\textit{Conclusion}}
The gallium anomaly has been an outstanding problem for about two decades. Resolutions involving vanilla eV-scale sterile neutrino run into strong tensions with solar and reactor measurements. In this work, we have studied tests for a new class of proposed solution based on fine-tuned resonant dips at energies of interest.  We propose detecting these low-energy neutrinos through the charged-current neutrino capture process on $^{115}$In. This can be readily realized by doping the opaque liquid organic scintillator experiment proposed by the LiquidO collaboration. We studied the prospects for such a detector combined with a 3.4 MCi $^{51}$Cr neutrino source. We also proposed to use either the 426\,keV line from the source or the $^7$Be solar neutrino lines to calibrate the neutrino capture cross section. The former case allows to provide stringent tests with ton-scale indium targets only in parts of the parameter space, whereas the latter requires larger detectors up 100\,ton indium mass, but can fully probe the parameter space.

Although we have only employed the MSW resonance model to test the sensitivity of the detector setup, many qualitative conclusions for the detector capabilities remain model agnostic and apply to sensitivity reach for other BSM interpretations of the GA anomaly including the vanilla eV-scale $\nu_s$ (see appendix for details).

The advent of Indium-doped LiquidO technology will help unlock new frontiers in sub-MeV neutrino detection by combining precision, scalability and strong background suppression. 

\label{sec:acknowledgements}
\textbf{\textit{Acknowledgements}}
We would like to thank Cecilia Lunardini for discussions on low-energy neutrino spectroscopy. The work is supported by the U.S. Department of Energy  Office of Science under the award number DE-SC0020262 (PH, GC) and NSF Award Number 2309973 (GC). GC also acknowledges the Center for Theoretical Underground Physics and Related Areas (CETUP*) and the Institute for Underground Science at SURF for hospitality during the completion of this work.

\bibliography{ref}

\clearpage
\appendix
\begin{widetext}
\section*{Appendices}

\section{Solar constraints from BOREXINO}
\begin{figure}[t!]
    \centering
    \includegraphics[width=0.6\linewidth]{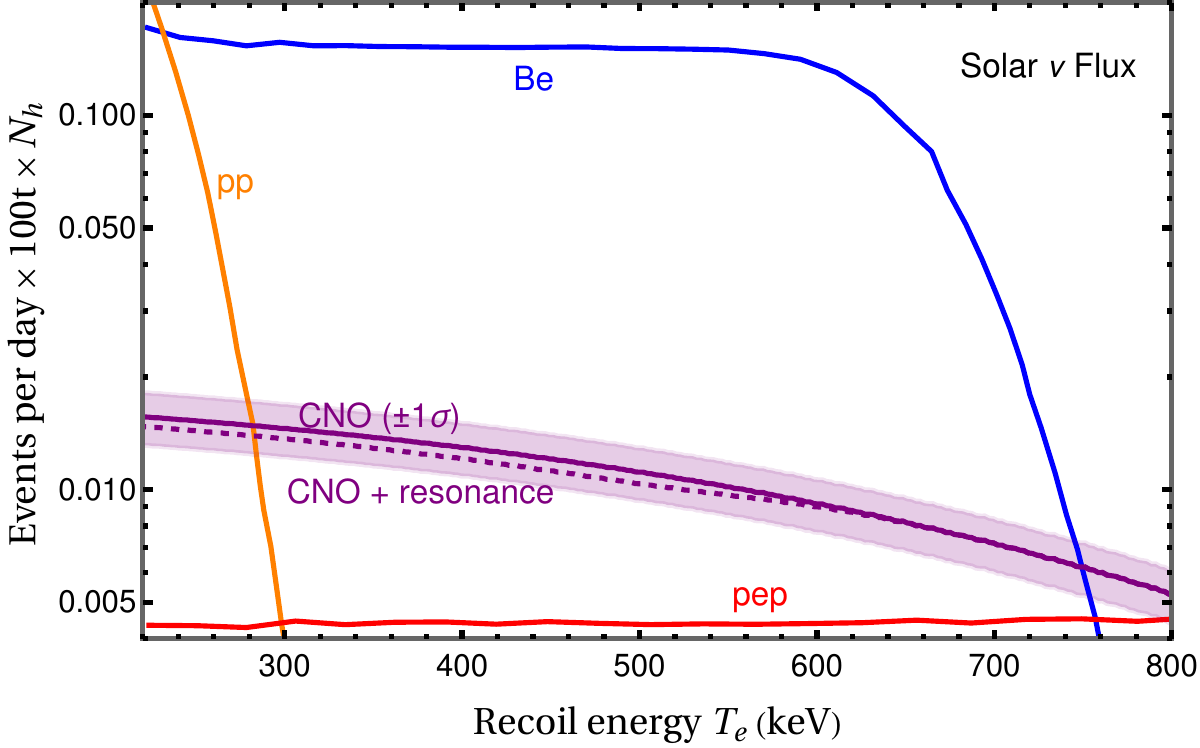}
    \caption{Electron recoil spectrum for various solar neutrino fluxes - ${}^{7}$Be, pp, pep and, CNO with and without the resonance in BOREXINO. The standard expected CNO events are shown by solid purple line with $1\sigma$ theoretical uncertainty region shown in shaded purple. The expected spectrum for a resonant $30\%$ dip centered at $750$ keV with width $50$ keV, is shown by dashed purple line.}
    \label{fig:p3}
\end{figure}
\label{sec:Borexino}
\noindent
The BOREXINO experiment has successfully measured multiple components of the solar neutrino flux~\cite{Bellini:2011rx,Borexino:2017rsf,BOREXINO:2018ohr}. Recently, the result of an improved and final measurement for the CNO  $\nu$-flux was reported~\cite{BOREXINO:2022abl,BOREXINO:2023ygs,BorexinoData}. Since the energies of interest for the GA overlaps with the CNO $\nu$-flux, the latest measurement can in principle be used to constrain the features of the MSW resonance. Such a resonance would lead to a sharp dip at the resonant energy in the neutrino flux and thus affecting the electron recoil spectrum event rate in the detector, which can be calculated as follows~\cite{Borexino:2019mhy}
\begin{equation}
    \frac{dR_\nu}{dT}= N_e \int dE_\nu \frac{d\Phi_\nu}{dE_\nu}\left[\frac{d\sigma_e}{dT}P_{ee}(E_\nu)+(\cos^2{\theta_{23}}\frac{d\sigma_\mu}{dT}+\sin^2{\theta_{23}}\frac{d\sigma_\tau}{dT})(1-P_{ee}(E_\nu)) \right]
\end{equation}
where $N_e$ is the number of $e^-$ in the fiducial volume of the detector, $d\Phi_\nu/dE_\nu$ is the differential neutrino flux spectrum, $P_{ee}(E_\nu)$ is the $\nu_e$ survival probability including the resonance effects, $d\sigma_\alpha(E_\nu,T)/dT$ is the differential neutrino-electron scattering cross-section w.r.t kinetic energy $T$ is given by
\begin{equation}
    \frac{d\sigma_\alpha(E_\nu,T)}{dT}= \frac{2}{\pi}G_F^2\,m_e\left[g_{\alpha L}^2 +g_{\alpha R}^2\left(1-\frac{T}{E_\nu}\right)^2 + g_{\alpha L} g_{\alpha R}\frac{m_e T}{E_\nu^2}  \right],\, \text{where } g_{\alpha L}=-\frac{1}{2}+\sin^2{\theta_W} + \delta_{\alpha e}, \, g_{\alpha R}=\sin^2{\theta_W} 
\end{equation}
with maximum recoil energy constrained to be 
\begin{equation}
    T_{\text{max}}=\frac{E_\nu}{1+\frac{m_e}{2E_\nu}}
\end{equation}

The electron recoil spectrum in the BOREXINO detector for relevant solar neutrino flux components - pp, pep, CNO and $^7$Be  - are shown in Fig.~\ref{fig:p3}. The solid lines denotes the measured recoil spectrum at BOREXINO along with $1\sigma$ theoretical uncertainty region shown in shaded purple, while the dotted line shows the modified recoil spectrum in presence of a resonant dip with $E_{\nu\text{,res}}=750$ keV, $P_{ee}(E_{\nu\text{,res}})\sim0.23$ (including solar oscillation effects) and resonance width $\sigma_{\text{res}}\simeq50$ keV. It can be clearly seen that the electron recoil spectrum cannot uniquely resolve the resonant dip in the CNO neutrino spectrum. The dip appears as a modified shoulder for $T_e$ below 560 keV (corresponding to $T_{\text{max}}$ for $E_{\nu\text{,res}}=750$ keV). For comparison, the benchmark point chosen by authors in ref.~\cite{Brdar:2023cms} corresponds to $E_{\nu\text{,res}}=750$ keV, $P_{ee}(E_{\nu\text{,res}})\sim0.28$ and $\sigma_{\text{res}}\simeq5$ keV. Since it is harder to resolve the sharp dip in the recoil spectrum even for a wider resonance width, therefore, it is not possible to test the viable parameter space for GA only through neutral scattering measurements. For this purpose, we argue for the need for a detector that can measure neutrino interactions through charged current reactions, to resolve the spectral features of the resonant dip and shed light on the nature of explanation that underlies gallium anomaly.

We also note that the measured CNO interaction rate is slightly higher than predicted in the HZ standard solar model (but still within $2\sigma$ C.I.)~\cite{Vinyoles:2016djt}. This might already disfavor resonant dips but since the measured rate by BOREXINO depends on the standard CNO spectral shape, this requires further work which is outside the scope of the current study.

\section{Neutrino capture cross-section}
\label{app:crosssection}
\noindent
For future studies, we provide the neutrino capture cross-section on ${}^{115}\text{In}$ in Fig.~\ref{fig:CrossSection} as a function of incoming neutrino energy $E_\nu$. The cross section for the neutrino capture by the Indium is~\cite{Konopinski:1959qr, Raghavan:1976yc}
\begin{equation}
    \sigma_\text{In}(E_{\nu})= \frac{2\pi^2 \ln{2}}{m_e^3}\frac{1}{(ft)_{{\text{inv} }\beta}} W_e\,p_e\,\textbf{F}(Z,W_e)
\end{equation}
where $(ft)_{{\text{inv} }\beta}=2.5\times 10^{4}$ s, $ W_e =\left(\frac{E_\nu-Q+m_e}{m_e}\right)$ is the total electron energy including the rest mass (in $m_e$ units), $p_e =\sqrt{W_e^2-1}$ is electron momentum (in $m_e$ units), $Q=114$ keV is the threshold energy and correction from the emitted electron in the nuclear Coulomb field $\textbf{F}(Z,W_e)$ is the Fermi function given by
\begin{align}
    \textbf{F}(Z,W_e) &= \,2(1+\gamma)\,(2p_e\,R)^{2\,(\gamma-1)}\,\times\text{Exp}\bigg(\pi\alpha\,Z\frac{W_e}{p_e}\bigg) \nonumber  \times \frac{|\Gamma(\gamma+i \alpha Z \frac{W_e}{p_e})|^2}{\Gamma(2\gamma+1)}
\end{align}
where $\gamma=(1-\alpha^2\,Z^2)^{1/2}$, $R = \frac{1}{2}\alpha\,A^{1/3}$, $\alpha$ is fine-structure constant, $Z=50$ is the atomic number and $A=115$ is the mass number. Note that we have adopted the same value of the $(ft)_{{\text{inv} }\beta}$ as argued and quoted in ref.~\cite{Raghavan:1976yc}. 
\begin{figure}[t!]
    \centering
    \includegraphics[width=0.5\linewidth]{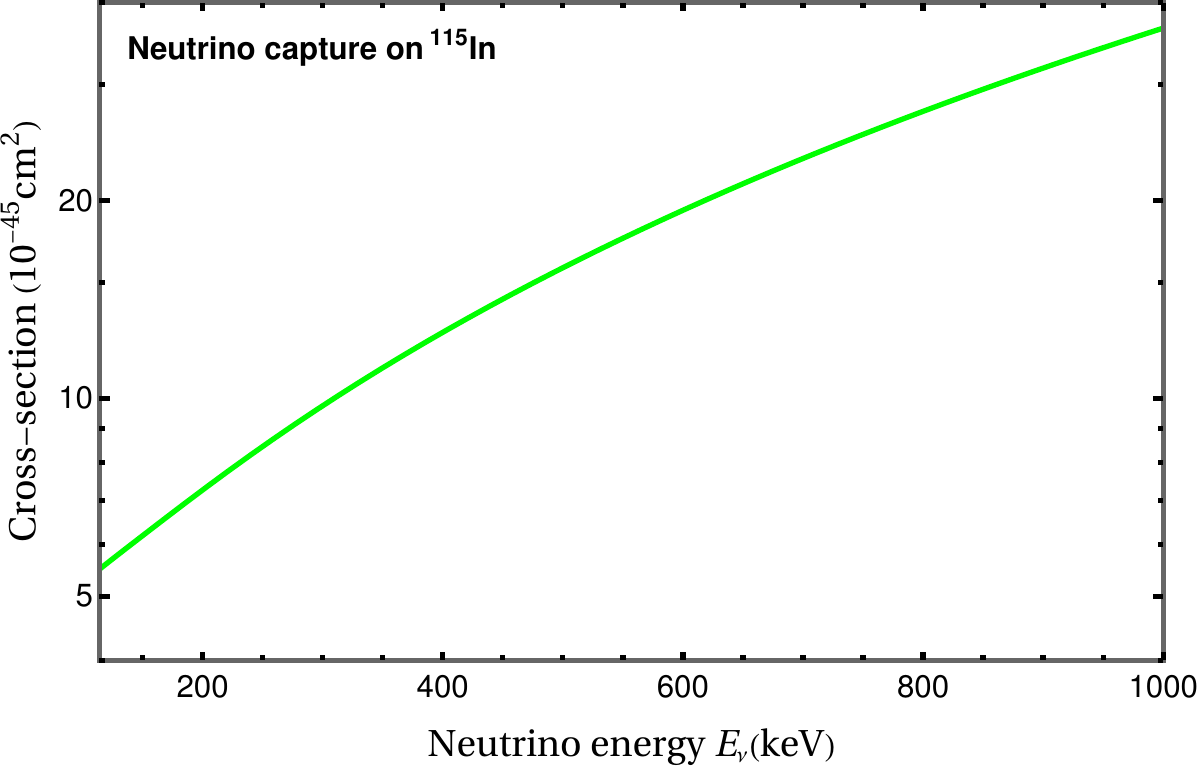}
    \caption{The cross-section for neutrino capture on ${}^{115}\text{In}$ (in units of $10^{-45}\text{ cm}^2$) as a function of neutrino energy $E_\nu$ (in keV).}
    \label{fig:CrossSection}
\end{figure}

\section{Event Rate Expressions}
\noindent
The differential electron recoil spectrum from neutrino capture in an In-doped LiquidO type detector with $^7$Be solar neutrino flux and $^{51}$Cr source respectively, is calculated as follows
\begin{equation}
    \left(\frac{d^2N_\nu}{dr\,dE_\text{R}}\right)_{{^{51}}\text{Cr}} = \frac{M_{\text{Ind}}}{A_{\text{Ind}}\,m_u\,V_\text{det}} \, \left( \frac{1-e^{-\lambda\,T_\text{obs}}}{\lambda} \right)\,\sum_{j=1}^{\text{all Cr $\nu$-lines}}\,N_{\nu,j}^\text{Cr}\,\sigma_\text{In}(E_{\nu,j})\,P_{ee}^{\text{MSW}}(E_{\nu,j},r)\, \mathcal{R}(E_{\nu,j},E_R)
\end{equation}
\begin{equation}
    \left(\frac{d^2N_\nu}{dr\,dE_\text{R}}\right)_{{^7}\text{Be}}= \frac{4\pi M_{\text{Ind}}}{A_{\text{Ind}}\,m_u\,V_\text{det}}\,T_{\text{obs}}\,r^2 \,\sum_{}^{\text{both $^7$Be $\nu$-lines}}\sigma_\text{In}(E_{\nu})\,P_{ee}^{\text{MSW}}(E_\nu)\,\Phi_{{^7}\text{Be}}({E_\nu})\, \mathcal{R}(E_{\nu},E_R)\, 
\end{equation}
where $\lambda=\ln{2}/T_{1/2}$, $T_{1/2}$ is the ${^{51}}\text{Cr}$ half-life, total observation time per source run $T_\text{obs}=200$ days, $M_{\text{Ind}}$ is the total Indium mass, $A_{\text{Ind}}=115$, $m_u$ is average nucleon mass, $V_{\text{det}}$ is the total detector mass, $N_{\nu,j}^\text{Cr}$ is the source neutrino rate for $E_{\nu,j}$ $\nu$-line, $\Phi_{{^7}\text{Be}}$ is ${^7}\text{Be}$ solar $\nu$-flux, $r$ is the radial distance from the center of the detector, $E_\nu$ is the incoming neutrino energy, $E_R$ is the reconstructed neutrino energy (electron recoil energy being $E_R-Q$), the energy reconstruction function $\mathcal{R}(E_{\nu},E_R)$ is assumed to be gaussian~\cite{Coloma:2022umy} with a width $\sigma_E=5\%/\sqrt{E_R(\text{in MeV})}$ (400 photoelectrons per MeV)~\cite{LiquidO:2019mxd}.

In fig.~\ref{fig:OscPee}, we show the oscillation probability $P_{ee}$ as a function of length $L$ and neutrino energy $E_\nu$ for 4 different benchmark point. Note that $B1$ is the benchmark point chosen in ref.~\cite{Brdar:2023cms} to explain GA. However, we find that while the energy dependence of the resonance at B1 agrees with the observed data, B1 lies significantly far away from the $3\sigma$ allowed region from the best-fit of all GA  experiments~\cite{Barinov:2022wfh}. We calculate and display the event rates in fig.~\ref{fig:eventsRates} for these 4 cases as a function of electron energy $(E_\nu-Q)$ and distance from $^{51}$Cr source placed inside an indium doped LiquidO setup. The above calculation assumes a single 3.4 MCi $^{51}$Cr source run of about 200 days.

\section{$\chi^2$-definition}
\noindent
We choose to define the $\chi^2$ in terms of number of events in each radial bin and for two energy bins for $E_\text{R}$ above and below $534$ keV, labeled $N_{i,750}^{\text{dip/SM}}$ and $N_{i,430}^{\text{dip/SM}}$ respectively

\begin{align*}
\chi^2(\vec{\xi}) = &\sum_{i=1}^{\text{all r bins}} 2\left[ 
\left(N_{i,430}^{\text{SM}}(\vec{\xi})-N_{i,430}^{\text{dip}}\right) + 
N_{i,430}^{\text{dip}}\ln\left(\frac{N_{i,430}^{\text{dip}}}{N_{i,430}^{\text{SM}}(\vec{\xi})}\right) + 
\left(N_{i,750}^{\text{SM}}(\vec{\xi})-N_{i,750}^{\text{dip}}\right) + 
N_{i,750}^{\text{dip}}\ln\left(\frac{N_{i,750}^{\text{dip}}}{N_{i,750}^{\text{SM}}(\vec{\xi})}\right)  
\right] \\
&+ \left( \frac{\xi_{\text{CrFlux}}^{430}}{\sigma_{\text{CrFlux}}^{430}} \right)^2 +
\left( \frac{\xi_{\text{BeFlux}}}{\sigma_{\text{BeFlux}}} \right)^2 +
\left( \frac{\xi_{\sigma\text{In}}}{\sigma_{\sigma\text{In}}} \right)^2 +
\left( \frac{\xi_{\text{BR}}}{\sigma_{\text{BR}}} \right)^2
\end{align*}

\begin{align*}
N_{i,430}^{\text{SM}}(\vec{\xi}) &= (1 +\xi_{\text{CrFlux}}^{430})(1 + \xi_{\sigma\text{In}}) \cdot { N_{i,\text{Cr,low}}^0} \\
&\simeq (1 + \xi_{\sigma\text{In}}+\xi_{\text{CrFlux}}^{430}) \cdot { N_{i,\text{Cr,low}}^0} \\
N_{i,750}^{\text{SM}}(\vec{\xi}) &= (1 + \xi_{\sigma\text{In}}) \left[
 \left(1- \frac{ \xi_{\text{BR}}B^0_L}{(1-B^0_L)}\right) \frac{(1 + \xi_{\text{CrFlux}}^{430})}{(1+\xi_{\text{BR}})}\cdot  N_{i,\text{Cr,high}}^0 +
(1 + \xi_{\text{BeFlux}}) \cdot N_{i,\text{Be7}}^0
\right] \\
&\simeq  \left(1- \frac{\xi_{\text{BR}}}{B^0_H}+ \xi_{\text{CrFlux}}^{430} + \xi_{\sigma\text{In}}\right)  \cdot  N_{i,\text{Cr,high}}^0 +
(1 + \xi_{\text{BeFlux}}+ \xi_{\sigma\text{In}}) \cdot N_{i,\text{Be7}}^0 
\end{align*}
where $N_{i,\text{Cr,high/low}}^0$ are the standard expected rates with branching ratios set to central values $B^0_{H/L}$. The ($\xi_{\text{CrFlux}},\xi_{\text{BeFlux}},\xi_{\sigma\text{In}},\xi_{\text{BR}}$) are the pull terms for $^{51}$Cr flux normalization at $430$ keV neutrino lines ($\sigma_{\text{CrFlux}}^{430}\sim 0.5\%$)~\cite{Veretenkin:2017kog,Gavrin:2021lam}, Be flux normalization ($\sigma_{\text{BeFlux}}\sim 2.6\%$)~\cite{BOREXINO:2018ohr}, Indium neutrino capture cross-section ($\sigma_{\sigma\text{In}}$) and the branching ratio for the Cr neutrino lines to $430$ keV lines ($\sigma_{\text{BR}}^{430}\sim 0.234\%$)~\cite{Fisher:1984rrq,KONSTANTINOV:1994200,YALCIN:2005,MOREIRA:2010}. Note that the calorimetric measurement only depends primarily on $320$ keV gamma ray, therefore it technically determines the total activity for the $427$ and $432$ keV neutrino lines within $\sigma_{\text{CrFlux}}^{430}\sim 0.5\%$. Here $N_{i}^{\text{SM}}$ is the expected event rate in the standard case, $N_{i}^{\text{dip}}$ is the event rate with the MSW resonance. We scan over a given Indium mass $M_{\text{Ind}}\simeq\{0.3,1.2,2,5,11,16,20,27,33,50,101\}$ tons assuming a $20\%$ loading by weight and obtain an optimum $M_{\text{Ind}}$ for each point in the parameter space for which the sensitivity exceeds $5\sigma$, with results shown in Fig.~\ref{fig:gridReach}.   

In fig.~\ref{fig:gridRes}, we uniformly scan over the allowed parameter space of the $\nu_s$-DM MSW resonance model and indicate the indium mass required to reach a confidence interval (C.I.) of greater than $5\sigma$ (colored points except black), with the In-doped LiquidO setup. The black points correspond to point in the parameter space for which the significance remains less than $5\sigma$ despite a large indium mass. The upper panels are results for single source runs with a floating precision on indium cross-section in the left panel and a $2.6\%$ precision in the right panel. The lower panels correspond to second and third source run for the $2.6\%$ precision case. It can be clearly seen that the $\nu_s$-DM MSW resonance model can be completely probed at $5\sigma$ level if the $\sigma_{\sigma\text{In}}\leq 2.6\%$.

For completeness, we similarly show the sensitivity of the above setup to the vanilla eV-scale sterile neutrino scenario in fig.~\ref{fig:gridVac}, although this model is highly disfavored from multiple different experiments (solar, reactor, cosmological). 

\begin{figure*}
    \centering
   {\includegraphics[width=0.49\linewidth]{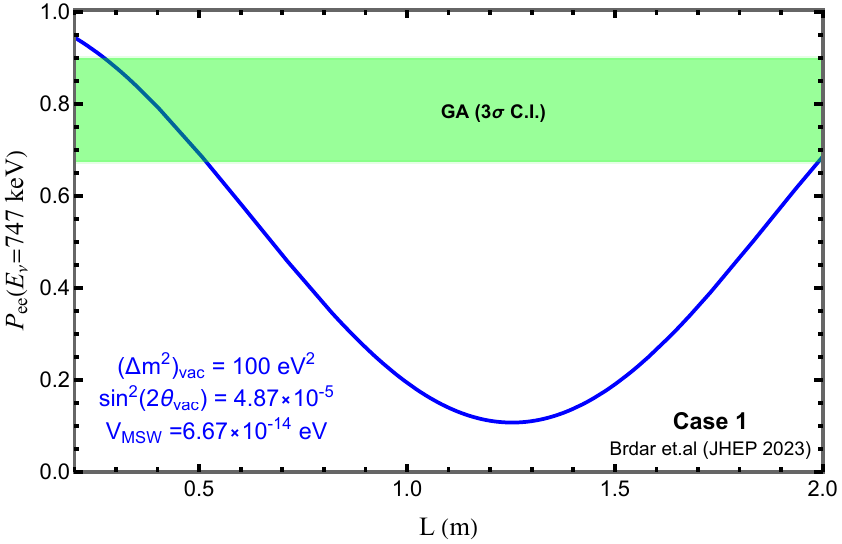}}
   {\includegraphics[width=0.49\linewidth]{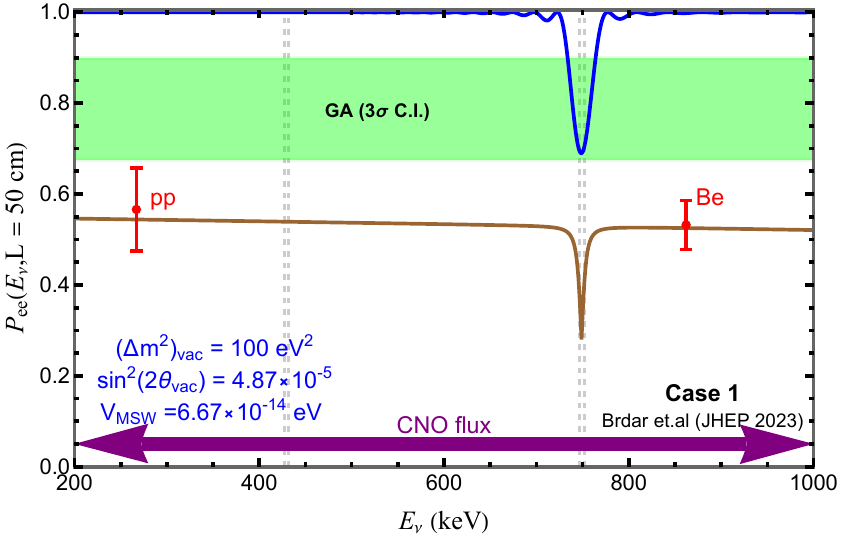}}
  {\includegraphics[width=0.49\linewidth]{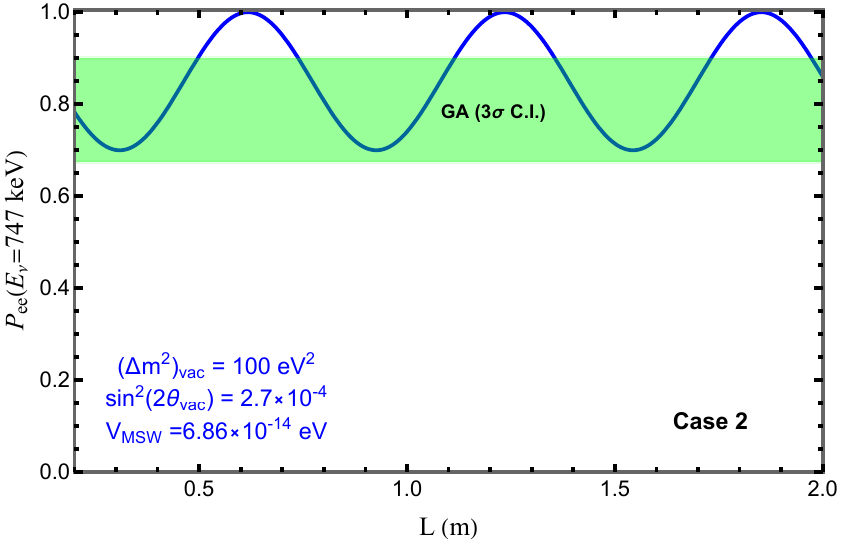}}
   {\includegraphics[width=0.49\linewidth]{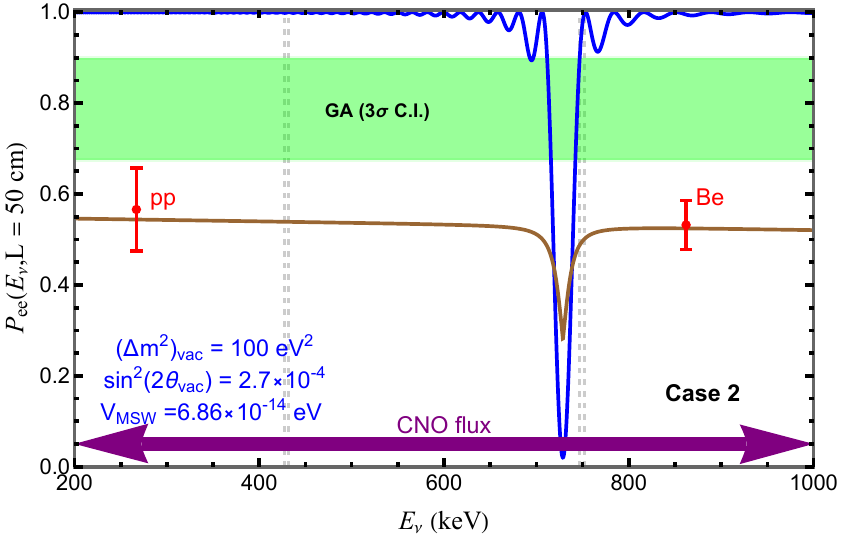}}
 {\includegraphics[width=0.49\linewidth]{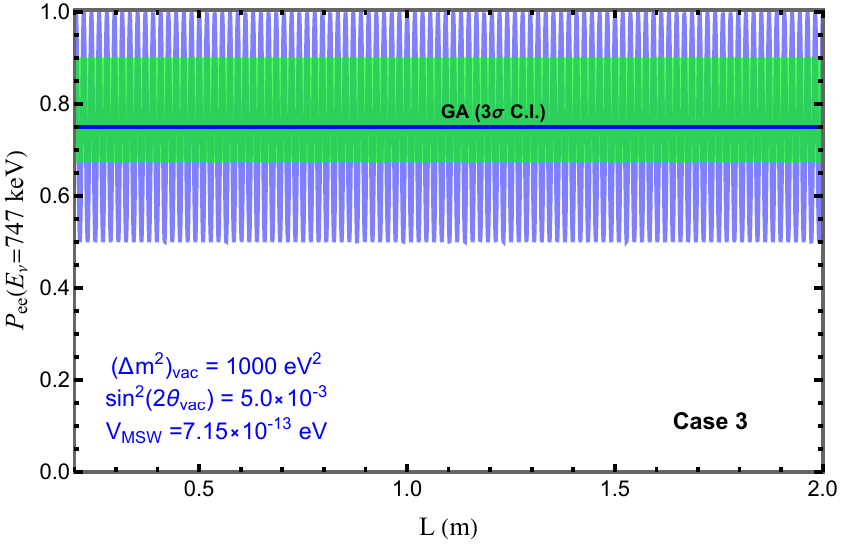}}
    {\includegraphics[width=0.49\linewidth]{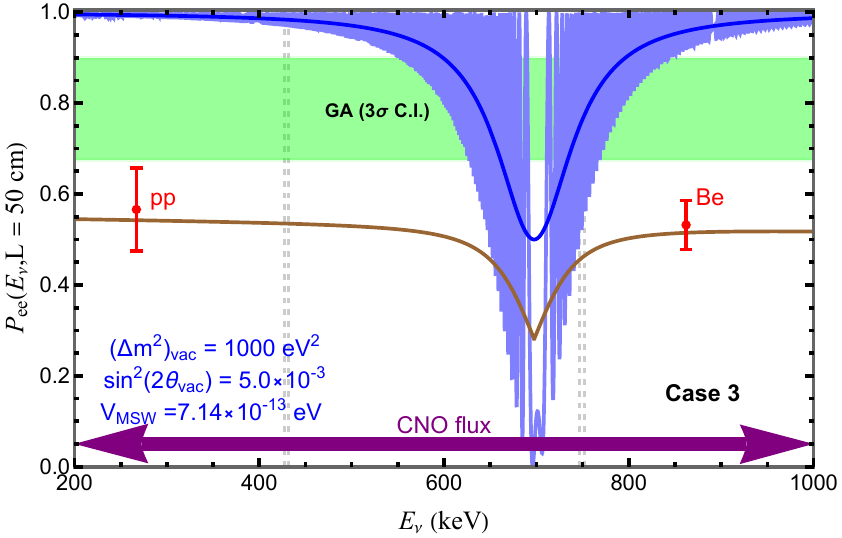}}
    {\includegraphics[width=0.49\linewidth]{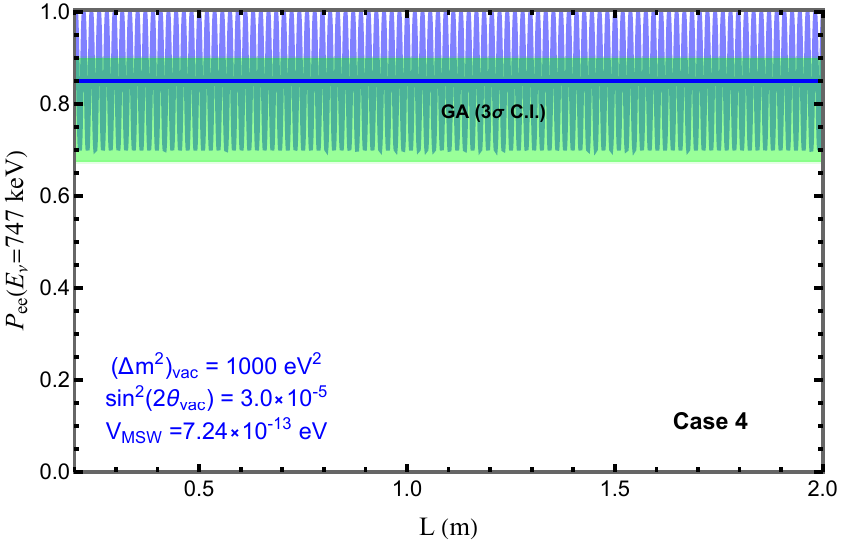}}
   {\includegraphics[width=0.49\linewidth]{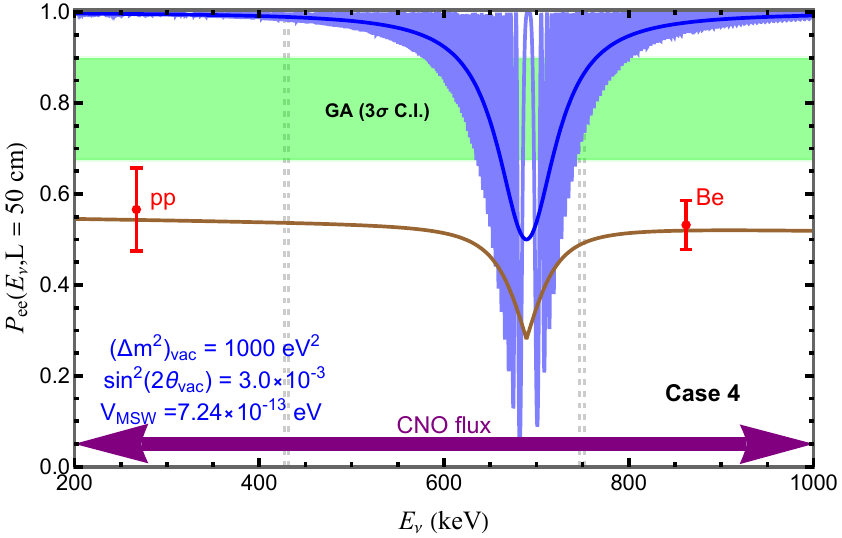}}
    \caption{The oscillation probability $P_{ee}$ as a function of length $L$ and neutrino energy $E_\nu$ for 4 different cases. Note B$1$ is the benchmark in ref.~\cite{Brdar:2023cms}. In effective parameter space, these cases correspond to $(\sin^2{2\theta^{\text{eff}}_{e4}},\Delta m^2_{\text{eff}})$ : B1\,$(0.89,0.74)$, B2\,$(0.3,3)$, B2\,$(0.5,100)$, B4\,$(0.3,100)$.}
    \label{fig:OscPee}
\end{figure*}

\begin{figure*}
    \centering
   {\includegraphics[width=0.49\linewidth]{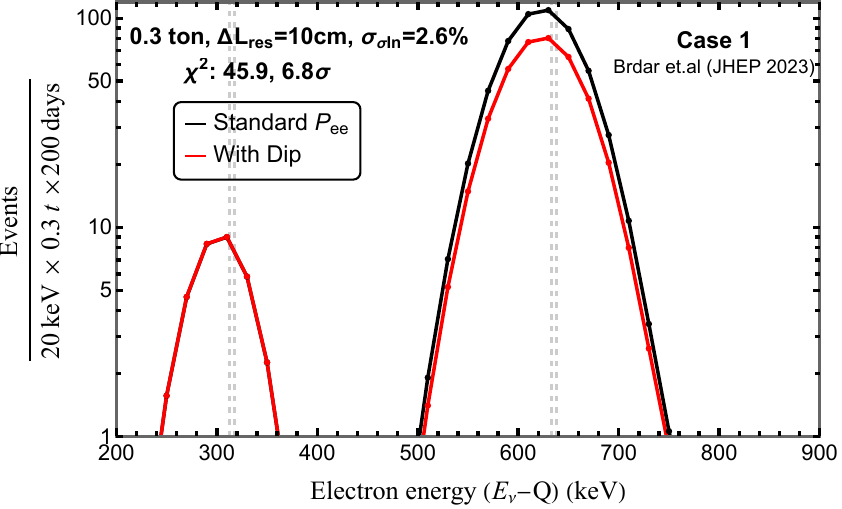}}
   {\includegraphics[width=0.49\linewidth]{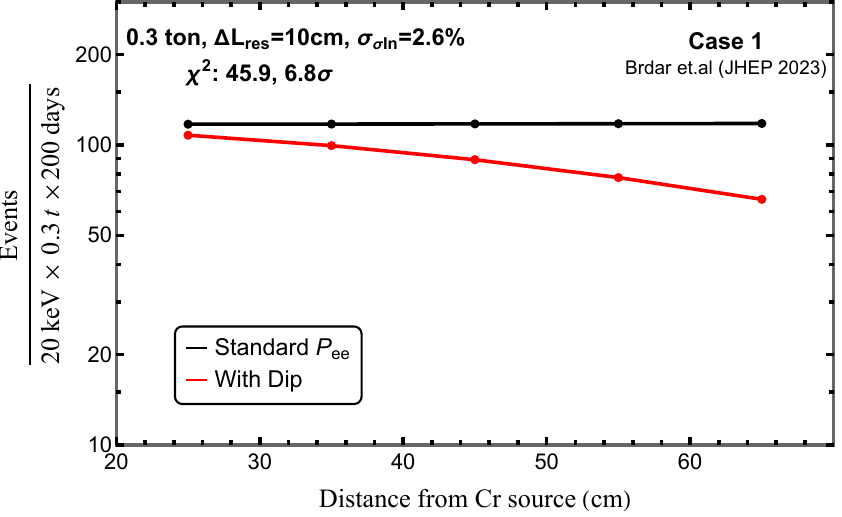}}
   {\includegraphics[width=0.49\linewidth]{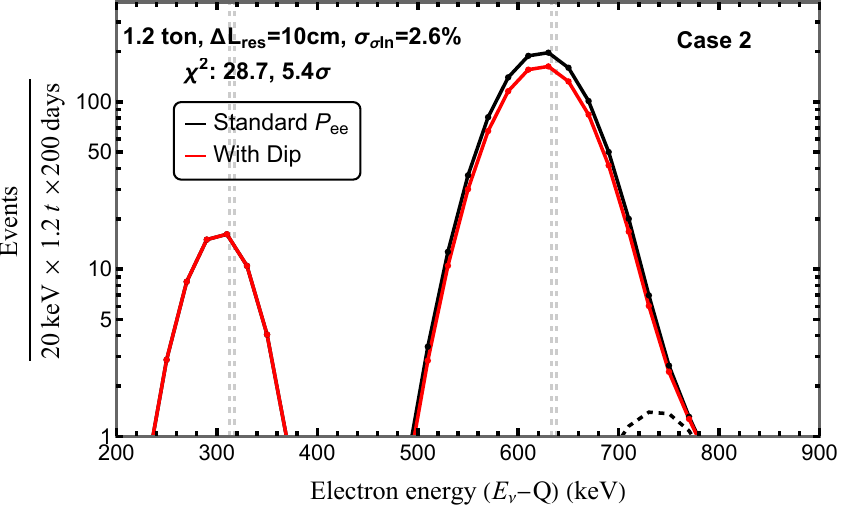}}
   {\includegraphics[width=0.49\linewidth]{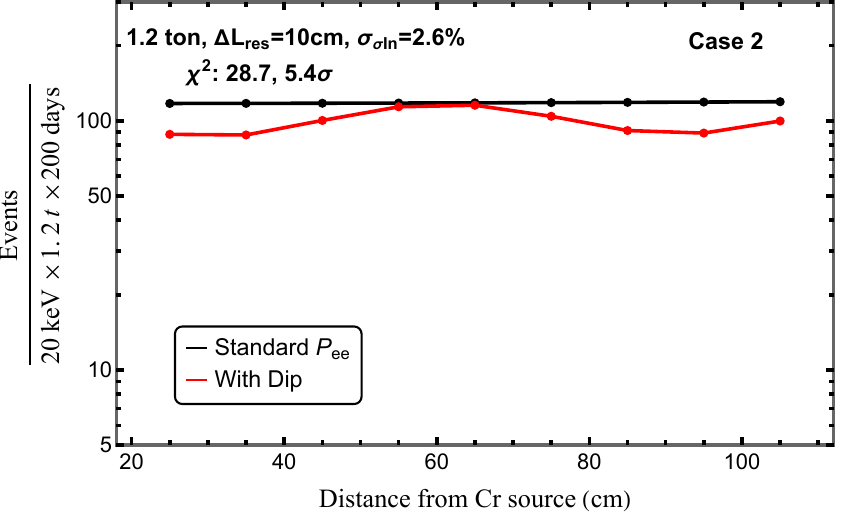}}
   {\includegraphics[width=0.49\linewidth]{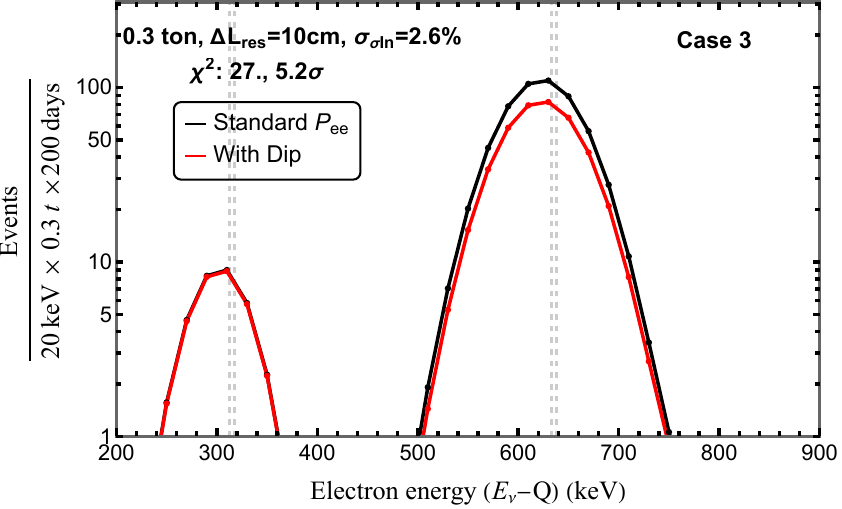}}
   {\includegraphics[width=0.49\linewidth]{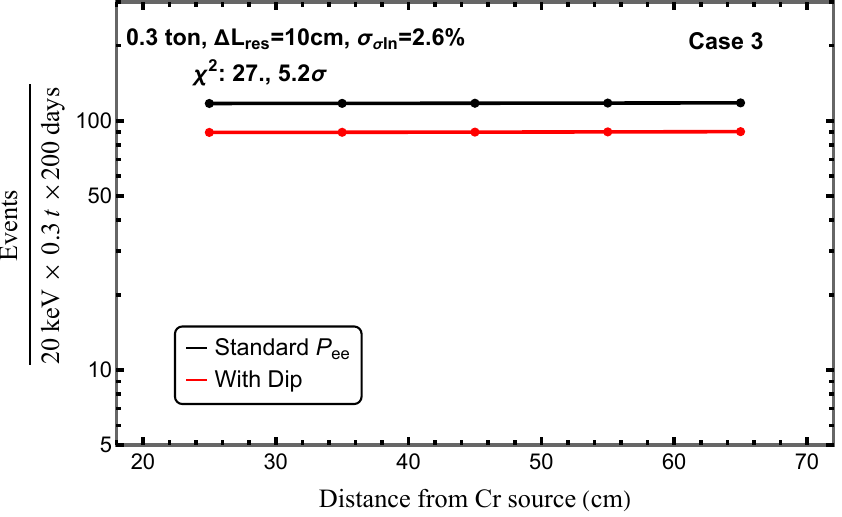}}
   {\includegraphics[width=0.49\linewidth]{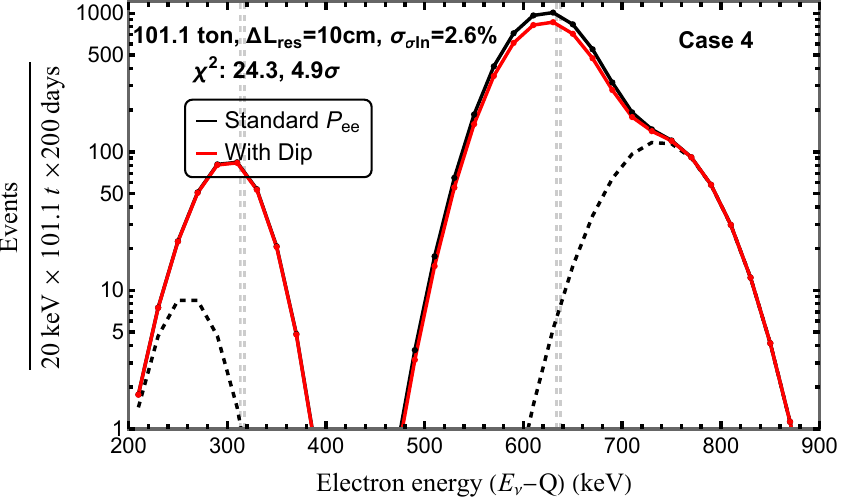}}
   {\includegraphics[width=0.49\linewidth]{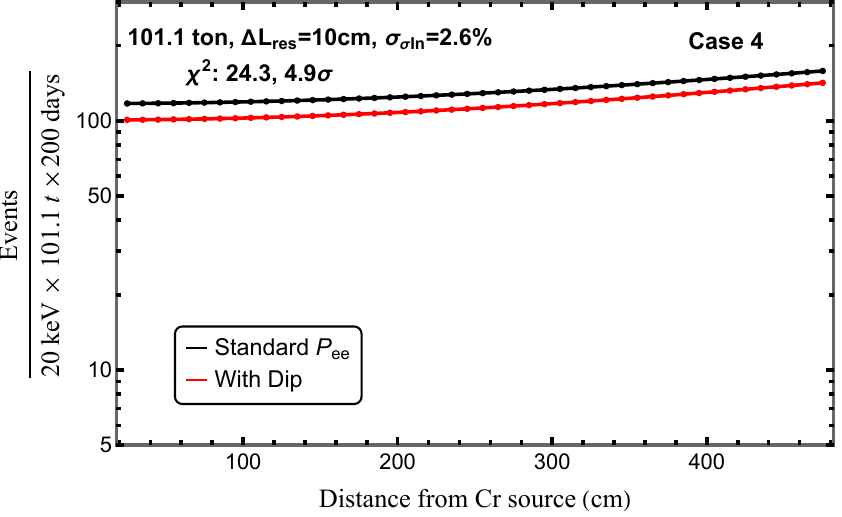}}
    \caption{Event rates as a function of electron energy $(E_\nu-Q)$ and distance from $^{51}$Cr source in an indium doped LiquidO style detector for a single source run of about 200 days, for the 4 benchmark points in Fig.~\ref{fig:OscPee}.}
    \label{fig:eventsRates}
\end{figure*}

\begin{figure*}
    \centering
    \includegraphics[width=0.495\linewidth]{plots/MainPlot_IndiumMass_Poisson_sigmIndFloating_SingleRun_Lres10cm_200days_Load20_3p4MCi_CrErr0p234_Mod2.pdf}
    \includegraphics[width=0.495\linewidth]{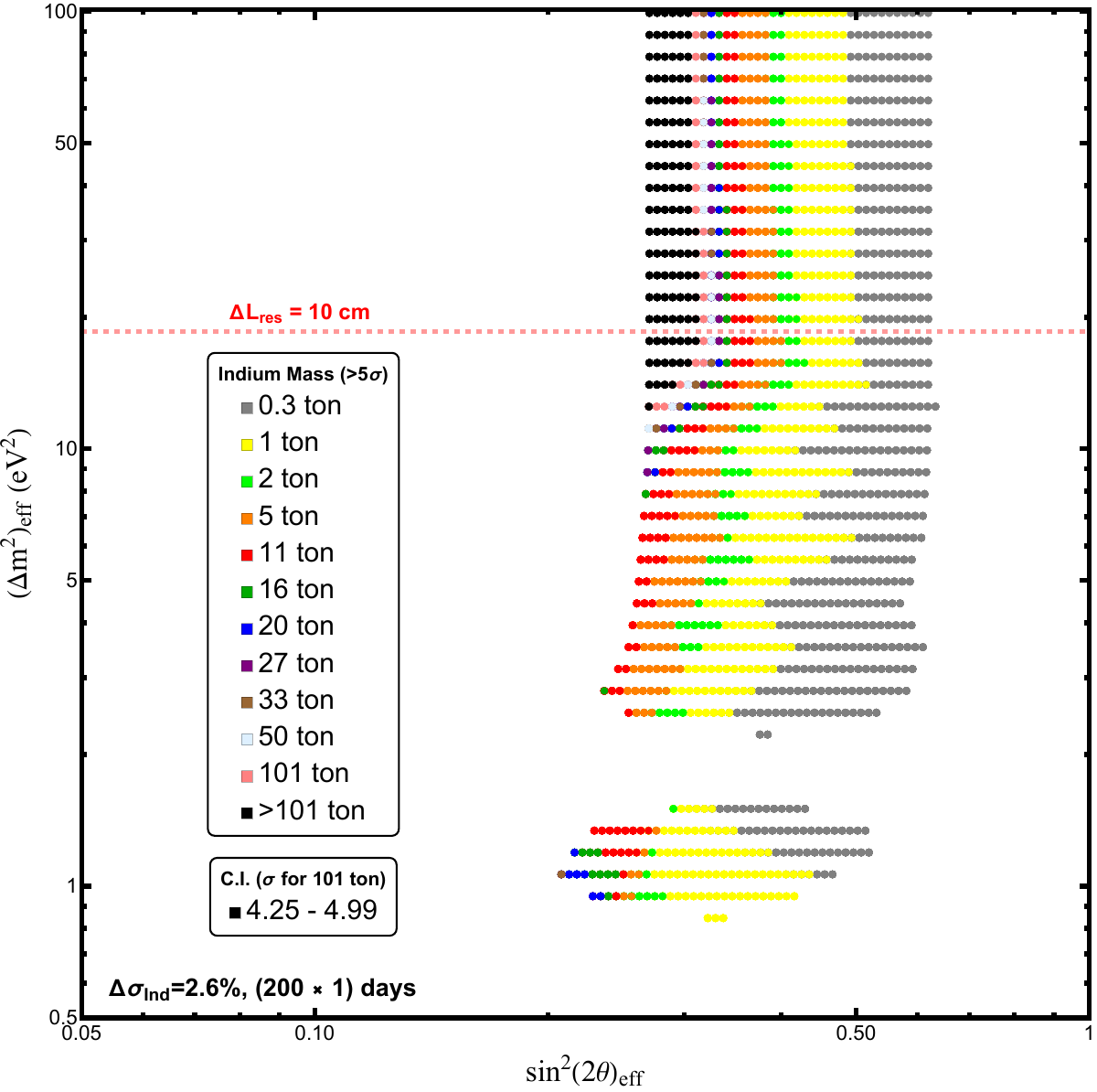}
    \includegraphics[width=0.495\linewidth]{plots/MainPlot_IndiumMass_Poisson_sigmInd2p6_DoubleRun_Lres10cm_400days_Load20_3p4MCi_CrErr0p234_Mod2.pdf}
    \includegraphics[width=0.495\linewidth]{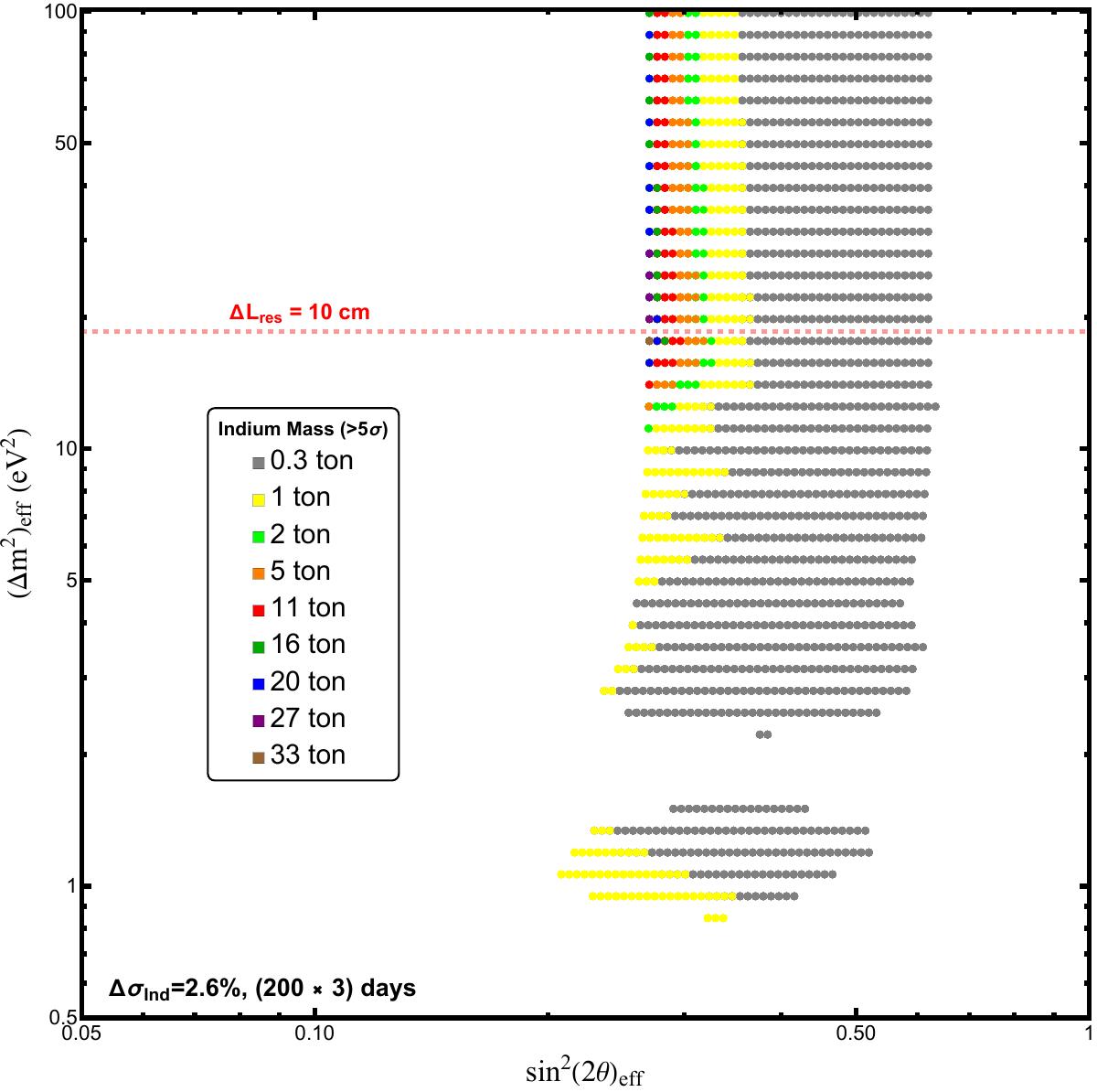}
    \caption{Plots for sensitivity of the In-doped LiquidO setup to effective parameter space of the $\nu_s$-DM MSW resonance model. The colored points indicate the indium mass required to reach a confidence interval (C.I.) of greater than $5\sigma$, except the black points for which the significance is less than $5\sigma$. All the points shown here are sampled uniformly from the allowed parameter space. The upper panels are results for single source runs with a floating precision on indium cross-section in the left panel and a $2.6\%$ precision in the right panel. The lower panels correspond to second and third source run for the $2.6\%$ precision case.}
    \label{fig:gridRes}
\end{figure*}

\begin{figure*}
    \centering
    \includegraphics[width=0.495\linewidth]{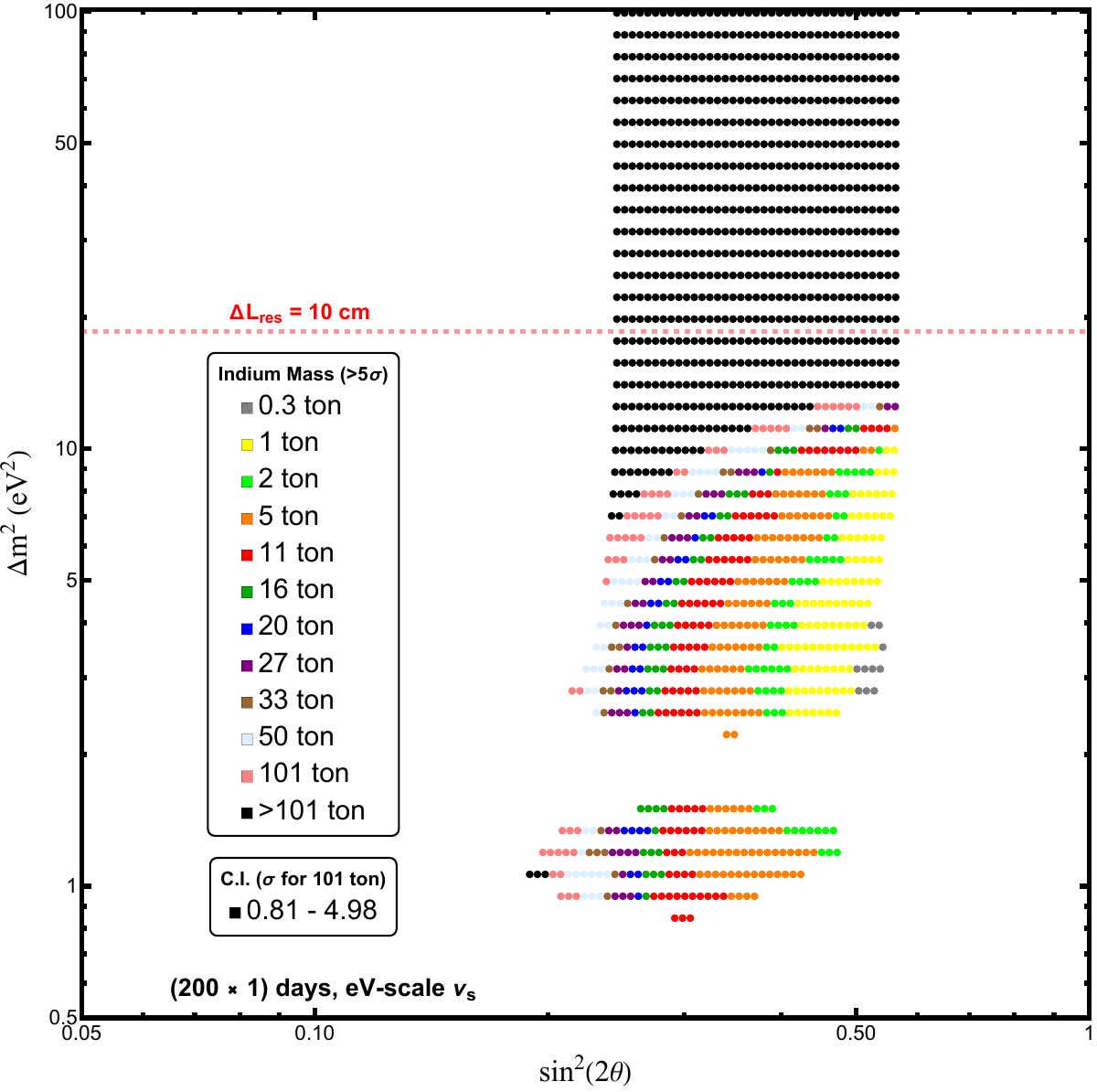}
    \includegraphics[width=0.495\linewidth]{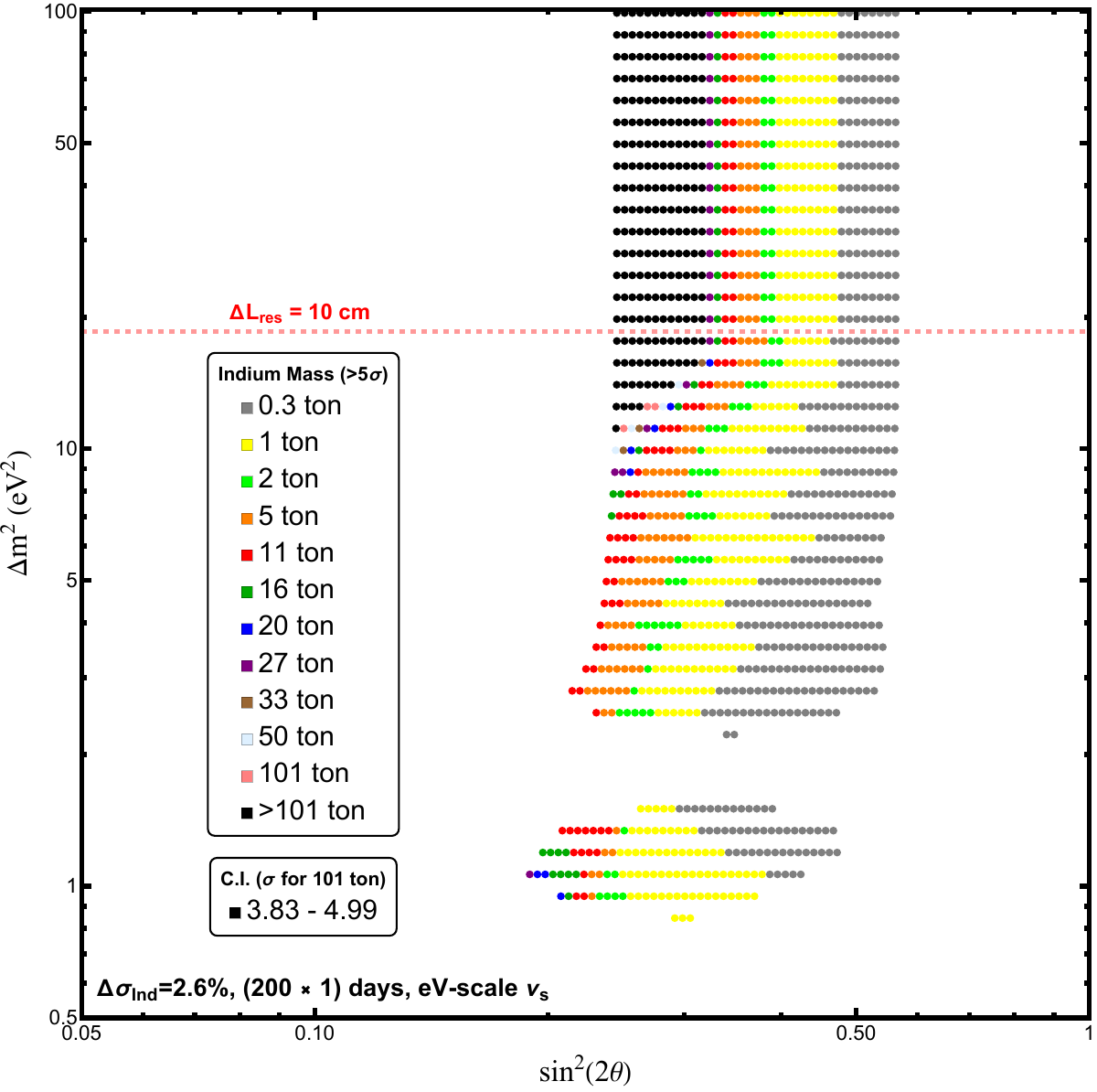}
    \includegraphics[width=0.495\linewidth]{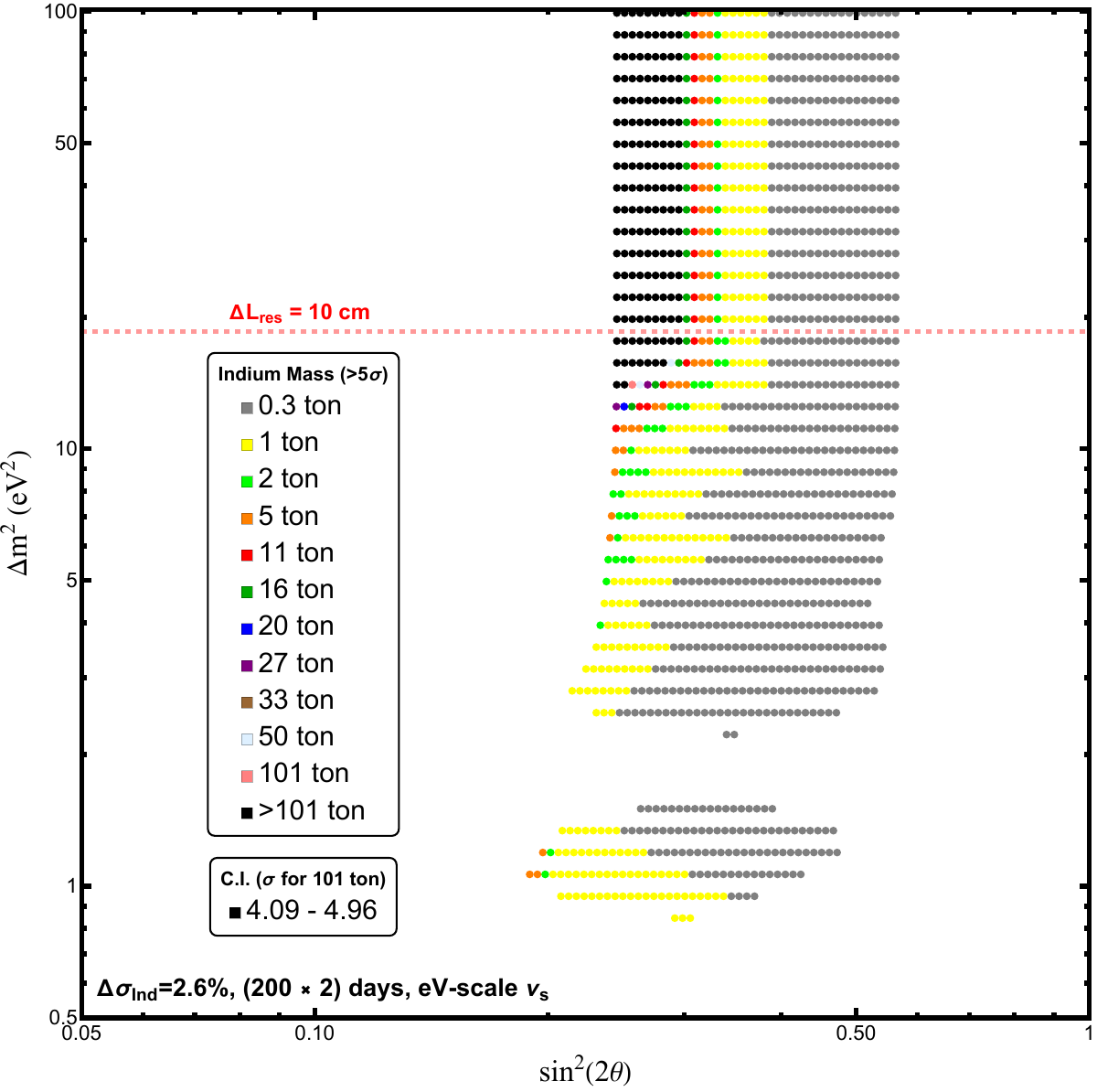}
    \includegraphics[width=0.495\linewidth]{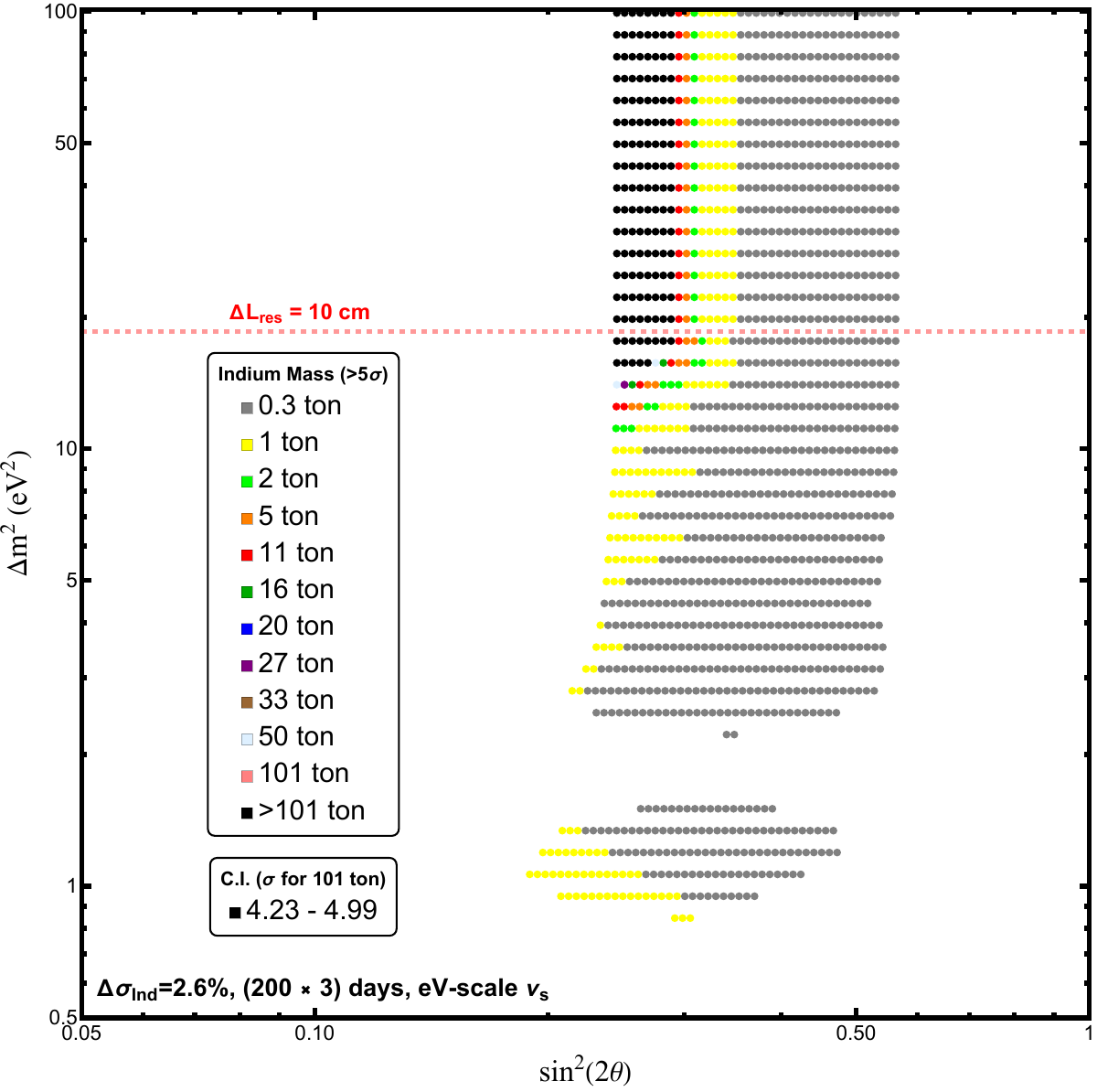}
    \caption{Plots for sensitivity of the In-doped LiquidO setup to the vanilla eV-scale sterile neutrino scenario. The colored points indicate the indium mass required to reach a confidence interval (C.I.) of greater than $5\sigma$, except the black points for which the significance is less than $5\sigma$. All the points shown here are sampled uniformly from the allowed parameter space. The upper panels are results for single source runs with a floating precision on indium cross-section in the left panel and a $2.6\%$ precision in the right panel. The lower panels correspond to second and third source run for the $2.6\%$ precision case. }
    \label{fig:gridVac}
\end{figure*}

\end{widetext}

\end{document}